 \documentstyle[sprocl,epsf,rotate]{article}
   
\def\Journal#1#2#3#4{{#1} {\bf #2}, #3 (#4)}

\def\ApJ{\em ApJ,}
\def\ApJL{\em ApJL,}
\def\AA{\em A\&A,}

\def\be{\begin{equation}}
\def\ee{\end{equation}}
\def\bea{\begin{eqnarray}}
\def\eea{\end{eqnarray}}

\def\etal{{\frenchspacing et al.}}
\def\syn{synchrotron~}

\def\microk{$\mu$K~}

\def\microm{$\mu$m~}

\def\rms{{\it rms~}}

\def\FigOne{
\begin{figure}
\makebox{
\noindent
\parbox[l]{3.0truecm}{\footnotesize
\caption{
Saskatoon and template maps.
For all maps, the temperatures are shown in coordinates where the
North Celestial Pole is at the center of a circle of 15$^{\circ}$
diameter, with RA=0 at the top and increasing clockwise.
The nine panels show the Saskatoon Ka Band map (Ka), the Saskatoon Q 
Band map (Q) and the full Saskatoon (Ka + Q Band) map (All), 
the 408 MHz Haslam survey (Ha), the 1420 MHz Reich \& Reich survey (RR), 
the point source template (PS), and the DIRBE 240, 140 and 100 \microm 
maps.
}
}
\hglue-0.08cm
\parbox[r]{9.0truecm}{
\rotate[r]{\vbox{\epsfxsize=8.8truecm\epsfbox{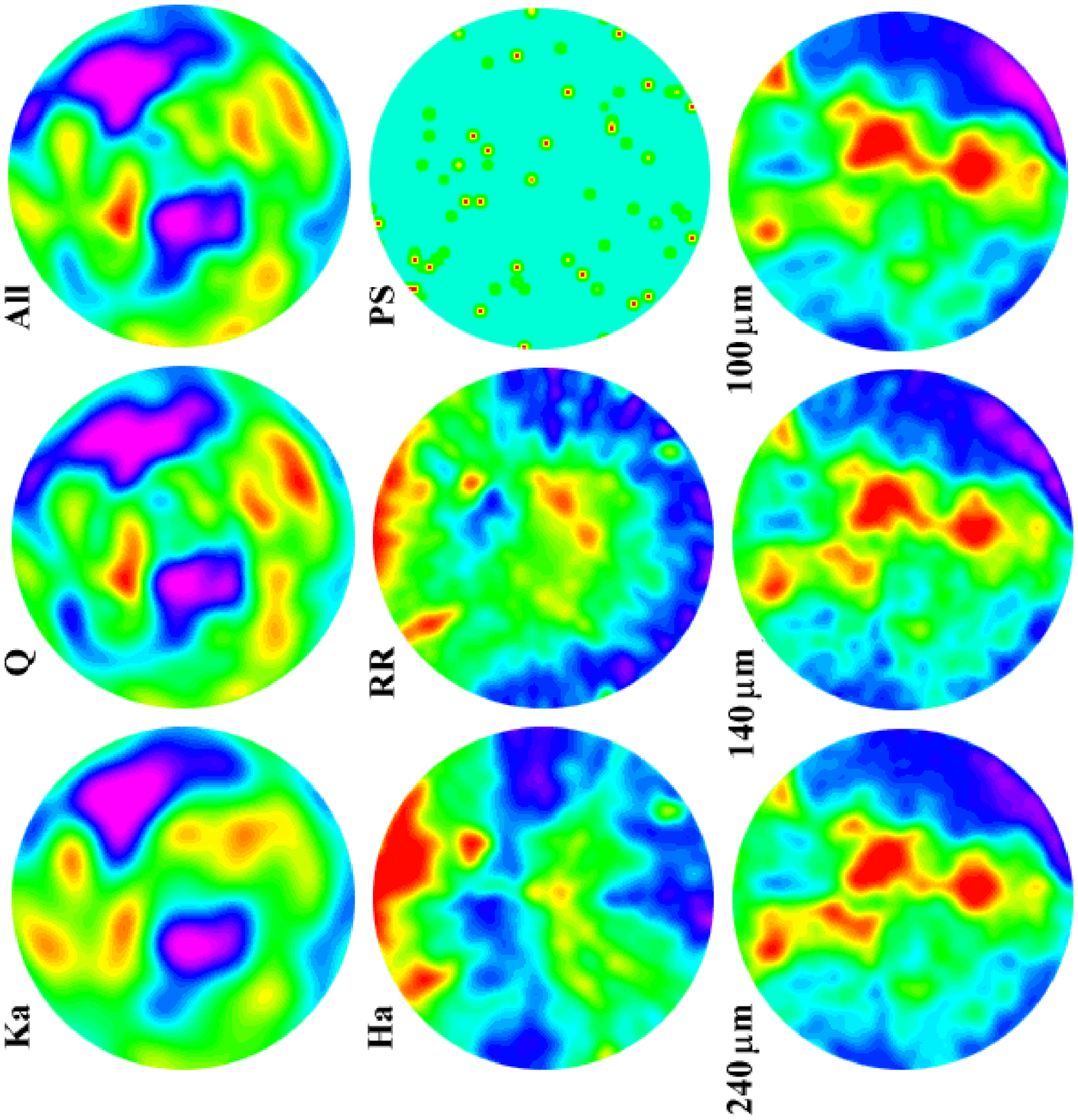}}}
\vskip-0.2truecm
}
}
\label{maps}
\end{figure}
}

\begin{document}

\title{FOREGROUND CONTAMINATION AROUND THE NORTH CELESTIAL POLE}
 
\author {ANG\'ELICA DE OLIVEIRA-COSTA}
\address{Princeton University, Department of Physics, Jadwin Hall, 
         Princeton, NJ 08544; angelica@pupggp.princeton.edu \\
         Institute for Advanced Study, Olden Lane, Princeton, 
         NJ 08540}
 
\author {AL KOGUT}
\address{Hughes STX Corporation, Laboratory for Astronomy and Solar
         Physics, Code 685, NASA/GSFC, Greenbelt MD 20771; 
         kogut@stars.gsfc.nasa.gov} 

\author {MARK J. DEVLIN, 
	 C. BARTH NETTERFIELD,
         LYMAN PAGE}
\author {EDWARD J. WOLLACK}
\address{Princeton University, Department of Physics, Jadwin Hall, 
         Princeton, NJ 08544; page@pupggp.princeton.edu} 

\maketitle\abstracts{
We cross-correlate the Saskatoon Q-Band data with different 
spatial template maps to quantify possible foreground contamination. 
We detect a correlation with the Diffuse Infrared Background 
Experiment (DIRBE) 100 \microm map, which we interpret as being 
due to Galactic free-free emission. Subtracting this foreground 
power reduces the Saskatoon normalization of the Cosmic Microwave 
Background (CMB) power spectrum by roughly 2\%. 
}

\section{INTRODUCTION}

One of the major concerns in any Cosmic Microwave Background 
(CMB) analysis is to determine if the observed signal is due to 
real CMB fluctuations or due to some foreground contaminant. 
At the frequency range and angular scale of the Saskatoon experiment 
\cite{Wollack,Netterfield}, there are two major potential 
sources of foreground contamination: diffuse Galactic emission and 
unresolved point sources.
The diffuse Galactic contamination includes three components: 
\syn and free-free radiation, and thermal emission from dust 
particles \cite{Partridge}. Although from a theoretical point of 
view, it is possible to distinguish these three components, there 
is no emission component for which both the frequency dependence and 
spatial template are currently well known \cite{Kogut_a}. 
The purpose of this paper is to use the Saskatoon data 
to estimate the Galactic emission at degree angular scales. 

\section{DATA ANALYSIS}

We based our analysis on the 1994-1995 data from Saskatoon 
experiment \cite{Wollack,Netterfield,Tegmark}. 
We cross-correlate the Saskatoon Q-Band data with two different \syn 
templates: the 408~MHz survey \cite{Haslam} and the 1420~MHz survey 
\cite{Reich}. To study dust and free-free emission, we cross-correlate 
the Saskatoon data with the Diffuse Infrared Background Experiment 
(DIRBE) sky map at wavelength 100~\microm \cite{Boggess}. In order to 
study the extent of point source contamination in the Saskatoon data, 
we cross-correlate it with the 1Jy catalog of point sources at 5~GHz 
\cite{Kuhr}. The templates used in this analysis, as well as the  
Saskatoon data, are shown in Figure~1.

\noindent\FigOne

\noindent 
The \syn templates, as well as the point source template, are found to 
be uncorrelated with the Saskatoon data. The DIRBE far-infrared template 
show a correlation, indicating a detection of signal with common spatial 
structure in the two data sets.
Kogut \etal \cite{Kogut_a,Kogut_b} detect a positive correlation 
between the DIRBE far-infrared maps and the DMR maps at 31.5, 53, and 
90~GHz, which they identify as being the result of a free-free emission. 
Assuming that this hypothesis can be extended to Saskatoon scales, we 
argue that the correlation between the DIRBE template and the Saskatoon 
data is most likely due to free-free contamination \cite{dOC}. 

\section{CONCLUSIONS}

In summary, we find a cross-correlation (at 97\% confidence) between the 
Saskatoon Q-Band data and the DIRBE 100 \microm map. 
The \rms amplitude of the contamination correlated with DIRBE
100 \microm is $\approx$ 17 \microk at 40 GHz.
We argue that the hypothesis of free-free contamination at degree 
angular scales is the most likely explanation for this correlated 
emission. Accordingly, the spatial correlation between dust and 
warm ionized gas observed on large angular scales seems to persist 
down to the smaller angular scales.

\bigskip
\noindent 
As reported by Netterfield \etal \cite{Netterfield}, the angular power 
spectrum from the Saskatoon data is $\delta T_{\ell}$=49$^{+8}_{-5}$ 
\microk at $l$=87 (corresponding to \rms fluctuations around 90 \microk 
on degree scales). This value of $\delta T_{\ell}$ is a much higher signal 
than any of the contributions from the foreground contaminants cited 
above, and shows that the Saskatoon data is not seriously contaminated by 
foreground sources. Since the foreground and the CMB 
signals add in quadrature, a foreground signal with 
	17\microk/90\microk $\approx$ 20\% 
of the CMB \rms only causes the CMB fluctuations to be over-estimated by 
$\sqrt{1+0.20^2} - 1 \approx 2\%$.

\end{document}